\documentclass[journal=jacsat,manuscript=article]{achemso}
\setkeys{acs}{articletitle = true}

\usepackage{amsmath,charter,bm,graphicx,multirow}
\usepackage[version=4]{mhchem}
\usepackage{booktabs}
\usepackage{multirow}
\usepackage[table,xcdraw]{xcolor}
\usepackage{soul}

\author{Juan\,R.\,Chamorro}
\affiliation{Materials Department and Materials Research Laboratory, University of California, Santa Barbara, California 93106, USA}
\email{jrchamorro@ucsb.edu}

\author{Julia\,L.\,Zuo}
\affiliation{Materials Department and Materials Research Laboratory, University of California, Santa Barbara, California 93106, USA}

\author{Euan\,N.\,Bassey}
\affiliation{Materials Department and Materials Research Laboratory, University of California, Santa Barbara, California 93106, USA}

\author{Aurland\,K.\,Watkins}
\affiliation{Materials Department and Materials Research Laboratory, University of California, Santa Barbara, California 93106, USA}

\author{Guomin\,Zhu}
\affiliation{Materials Department and Materials Research Laboratory, University of California, Santa Barbara, California 93106, USA}

\author{Arava\,Zohar}
\affiliation{Materials Department and Materials Research Laboratory, University of California, Santa Barbara, California 93106, USA}

\author{Kira\,E.\,Wyckoff}
\affiliation{Materials Department and Materials Research Laboratory, University of California, Santa Barbara, California 93106, USA}

\author{Tiffany\,L.\,Kinnibrugh}
\affiliation{X-ray Science Division, Advanced Photon Source, Argonne National Laboratory, 9700 S. Cass Ave, Argonne, Illinois 60439, USA}

\author{Saul\,H.\,Lapidus}
\affiliation{X-ray Science Division, Advanced Photon Source, Argonne National Laboratory, 9700 S. Cass Ave, Argonne, Illinois 60439, USA}

\author{Susanne\,Stemmer}
\affiliation{Materials Department and Materials Research Laboratory, University of California, Santa Barbara, California 93106, USA}

\author{Rapha\"ele\,J.\,Cl\'ement}
\affiliation{Materials Department and Materials Research Laboratory, University of California, Santa Barbara, California 93106, USA}

\author{Stephen\,D.\,Wilson}
\affiliation{Materials Department and Materials Research Laboratory, University of California, Santa Barbara, California 93106, USA}

\author{Ram\,Seshadri}
\affiliation{Materials Department and Materials Research Laboratory, University of California, Santa Barbara, California 93106, USA}
\email{seshadri@mrl.ucsb.edu}

\title[charter]{Soft-Chemical Synthesis, Structure Evolution, and Insulator-to-Metal Transition in a Prototypical Metal Oxide, $\lambda$-RhO$_2$}

\begin{document}

\clearpage

\begin{abstract}
 
$\lambda$-RhO$_2$, a prototype 4\textit{d} transition metal oxide, has been prepared by oxidative delithiation of 
spinel LiRh$_2$O$_4$ using ceric ammonium nitrate. Average-structure studies of this RhO$_2$ polytype, including synchrotron powder X-ray diffraction and electron diffraction, indicate the room temperature structure to be tetragonal, in the space group $I4_1/amd$, with a first-order structural transition to cubic $Fd\bar3m$ at $T$\,=\,345\,K on warming. Synchrotron X-ray pair distribution function analysis and $^7$Li solid state nuclear magnetic resonance measurements suggest that the room temperature structure displays local Rh--Rh bonding. The formation of these local dimers appears to be associated with a metal-to-insulator transition with a non-magnetic ground state, as also supported by density functional theory-based electronic structure calculations. This contribution demonstrates the power of soft chemistry to kinetically stabilize a surprisingly simple binary oxide compound. 
\end{abstract}

\clearpage

\section{Introduction}

In the presence of strong spin-orbit coupling, as expected in 4\textit{d} and 5\textit{d} transition metals, a \textit{d}$^5$ electronic configuration can give rise to an effective \textit{j} = 1/2 ground state with enhanced correlations, paving the way for exotic long-range macroscopic quantum states such as in $\alpha$-RuCl$_3$ \cite{Plumb2014, Banerjee2017, Banerjee2018, Do2017, Yadav2016}, Sr$_2$IrO$_4$ \cite{Kim2008, Kim2009, Kim2012, Zwartsenberg2020}, Sr$_3$Ir$_2$O$_7$ \cite{Nagai2007, Boseggia2012, Mazzone2022}, and \textit{A}$_2$IrO$_3$ (\textit{A}\,= Li, Na) \cite{ Biffin2014, Takayama2015, Williams2016, Halloran2022, Singh2010,Liu2011,Choi2012, Comin2012}. Compounds containing tetravalent rhodium (Rh$^{4+}$, 4\textit{d}$^5$), which is isoelectronic to Ru$^{3+}$ and 
is the 4\textit{d} analogue of Ir$^{4+}$, should possess electronic and magnetic properties similar to those observed 
in other \textit{d}$^5$ compounds. However, studies of the chemistry and physics of Rh$^{4+}$ in the solid state phase space remain rather limited due to the relatively greater stability of the trivalent 
state (Rh$^{3+}$, 4\textit{d}$^6$). In RhO$_6$ octahedra (the most common coordination environment in rhodium oxides, owing 
to cation size effects), Rh$^{3+}$ assumes the low-spin \textit{d}$^6$ configuration, with fully filled \textit{t}$_{2g}$ 
states. This stable electronic configuration results in an enhancement of the Rh-O bond strength that is comparatively 
destabilized in Rh$^{4+}$, which instead possesses a hole in the low-energy \textit{t}$_{2g}$ manifold. 

A search of the Inorganic Crystal Structure Database (ICSD) yields a total of around 266 oxide compounds containing 
Rh$^{3+}$, versus 30 unique oxides containing Rh$^{4+}$. The difficulty in synthesizing oxide compounds with Rh$^{4+}$ 
originates from the high oxidative potential required to oxidize Rh$^{3+}$ to Rh$^{4+}$, as well as the 
predilection for RhO$_2$ to vaporize\cite{Alcock1960, Jacob2010}. These issues have been addressed by performing 
syntheses under high oxygen pressures that can stabilize higher oxidation states and arrest volatilization, and have yielded a number of rhodium(IV) oxides such as \textit{A}RhO$_3$ (\textit{A} = Ca, Sr, and Ba) \cite{Yamaura2009,Li2017,Chamberland1981}, Sr$_3$Rh$_2$O$_7$ \cite{Yamaura2002}, 
and Sr$_4$Rh$_3$O$_{10}$ \cite{Yamaura2004}. In fact, the synthesis of RhO$_2$ in either its rutile \cite{Shannon1968} 
or cubic \cite{Shirako2014} forms requires high oxygen pressure in order to form.

Of the few reported Rh$^{4+}$ oxide compounds in the literature, only a small number have been demonstrated to possess phenomena similar to other aforementioned \textit{d}$^5$ materials, such as in the correlated electron metal Sr$_2$RhO$_4$ \cite{Shimura1992,Perry2006,Battisti2020}, spin-glass Li$_2$RhO$_3$ \cite{Todorova2010,Luo2013,Khuntia2017}, and the mixed-valent, Rh$^{3+}$/Rh$^{4+}$ spinel LiRh$_2$O$_4$ \cite{Okamoto2008,Knox2013,Shiomi2022}. In contrast to other \textit{d}$^5$ systems, however, spin-orbit coupling appears to play a near negligible role in establishing the ground state of Rh$^{4+}$ oxides, and other single-ion instabilities, such as Jahn-Teller distortions, often play a much larger part. This is the case for mixed-valent LiRh$_2$O$_4$, which crystallizes at room temperature in the prototypical cubic \textit{Fd}$\bar{3}$\textit{m} spinel structure, but distorts at low temperature into a tetragonal cell owing to a Jahn-Teller instability of the Rh$^{4+}$ octahedra, and then to an orthorhombic cell on further cooling due to charge-ordering \cite{Knox2013,Shiomi2022}. While many rhodium oxide spinels exist, 
including CoRh$_2$O$_4$ \cite{Cascales1984,Ge2017}, NiRh$_2$O$_4$ \cite{Horiuti1964,Chamorro2018}, and 
CuRh$_2$O$_4$ \cite{Ge2017,Dollase1997}, LiRh$_2$O$_4$ is the only one possessing any Rh$^{4+}$.

The Rh cations in LiRh$_2$O$_4$ form a three-dimensional pyrochlore network, and thus interactions between them can be destabilized by geometric frustration. A total topotactic delithiation of LiRh$_2$O$_4$ to Rh$_2$O$_4$ would result in a pyrochlore network of exclusively Rh$^{4+}$, which would be a potential \textit{j} = 1/2 Rh analog of the \textit{RE}$_2$Ir$_2$O$_7$ (\textit{RE} = Y, Pr$-$Lu) pyrochlore iridates. Some of these pyrochlore iridates display metal-to-insulator transitions \cite{Matsuhira2011,WitczakKrempa2014,Wan2011} arising from strong interactions between Ir cations, and possess non-trivial band topologies that give rise to various exotic states such as Weyl semimetal \cite{Yang2011,Sushkov2015,Li2021,Liu2021} and Luttinger liquid \cite{Ohtsuki2019,Mandal2021,arXivNikolic2022} states. In order to synthesize Rh$_2$O$_4$, we sought to topotactically remove Li$^{+}$ cations from LiRh$_2$O$_4$ using electrochemical cells and chemical agents. Initial tests using these cells and common solution-based delithiating oxidants such as Br$_2$ and I$_2$ in acetonitrile did not reveal obvious changes to the crystallographic structure, as determined by X-ray diffraction (XRD). We therefore turned to other chemical oxidants with oxidation potentials greater than those of Br$_2$ and I$_2$, in order to overcome the aforementioned Rh$^{3+}\rightarrow$\,Rh$^{4+}$ redox barrier.

In this work, we report the topotactic, oxidative delithiation of LiRh$_2$O$_4$ to form a new rhodium (IV) oxide, $\lambda$-RhO$_2$. Fashioned after $\lambda$-MnO$_2$, which was also obtained via soft, chemical delithiation of LiMn$_2$O$_4$ spinel in acid\cite{Hunter1981}, we have employed the use of ceric ammonium nitrate (NH$_4$)$_2$Ce(NO$_3$)$_6$ to remove nearly all of the lithium cations in LiRh$_2$O$_4$ topotactically, retaining the parent spinel architecture. $^7$Li solid-state nuclear magnetic resonance (NMR) measurements indicate that the nominal lithium content is 84(1)\% reduced as compared to the parent compound. To our knowledge, this is the first reported application of ceric ammonium nitrate, a powerful shelf-stable oxidizer with an oxidizing potential superior to that of elemental chlorine, for the topotactic oxidative delithiation of an extended solid. Our results indicate that $\lambda$-RhO$_2$ is a metal at high temperatures in excess of \textit{T} = 350 K and crystallizes in the cubic \textit{Fd}$\bar{3}$\textit{m} space group, while it undergoes a hysteretic metal-to-insulator transition on cooling that reduces the average structure to tetragonal \textit{I}\textit{4}$_1$/\textit{amd}. This transition is accompanied by the formation of short-range Rh-Rh dimers, formed through direct metal-metal bonding, and results in a non-magnetic ground state. This work expands the \textit{A}Rh$_2$O$_4$ rhodium oxide spinel phase space to the extreme limit where \textit{A} is occupied by a vacancy.

\section{Results and Discussion}

\subsection{Topotactic Oxidative Delithiation Reactions}

Oxidative delithiation of LiRh$_2$O$_4$ was performed using ceric ammonium nitrate (CAN), a reagent commonly used in other fields of synthetic and industrial chemistry \cite{Nair2007,Molander1992,Xiao2000,Hwu2010}, but rarely used in synthetic solid state chemistry. It is a powerful, one electron oxidizing agent that is driven by the single electron process Ce$^{4+}$ + \textit{e}$^{-}$ $\rightarrow$ Ce$^{3+}$, which has a reduction potential of \textit{E}$^{\circ}$ = 1.61 V \cite{Nair2007}. This potential is higher than that of other commonly used powerful oxidizers for the topotactic deintercalation of extended solids, such as Cl$_2$ and Br$_2$ (\textit{E}$^{\circ}$ = 1.36 and 1.08 V, respectively) \cite{Vanysek2005}, and is on par with permanganates (MnO$_4$$^-$, 1.51$-$1.67 V). CAN is a non-hygroscopic, shelf-stable compound that can be readily handled in air and is soluble in aqueous and organic solvents. Its use in low temperature \textit{chimie douce} reactions remains largely unexplored in the synthetic materials chemistry literature, and our findings present a case where CAN can be employed to stabilize a rare, tetravalent oxidation state in rhodium.

This article is focused primarily on the nearly-fully-delithiated end member of the Li$_{1-x}$Rh$_2$O$_4$ phase space, $\lambda$-RhO$_2$ (Li$_{0.1(1)}$Rh$_2$O$_4$). This compound is synthesized using an excess amount of CAN in water, whereas other Li$_{1-x}$Rh$_2$O$_4$ compounds are synthesized by selecting a target CAN concentration in each solution per mol of LiRh$_2$O$_4$. The proposed chemical equation for the oxidative delithiation reaction is: 

\begin{center}
    LiRh$_2$O$_4$ + \textit{x}(NH$_4$)$_2$Ce$^{\mathrm{IV}}$(NO$_3$)$_6$ $\rightarrow$ Li$_{1-x}$Rh$_2$O$_4$ + \textit{x}(NH$_4$)$_2$Ce$^{\mathrm{III}}$(NO$_3$)$_5$ + \textit{x}LiNO$_3$
\end{center}

Based on this reaction, one mole of CAN is needed to fully delithiate LiRh$_2$O$_4$. We prepared samples of Li$_{1-x}$Rh$_2$O$_4$ at \textit{x} = 0.1 intervals, and performed synchrotron X-ray diffraction measurements at \textit{T} = 400 K, as discussed in more depth further in this text, in order to track their lattice parameters as a function of targeted delithiation. At this temperature, every sample had a cubic structure, and the cubic lattice parameter as a function of the relative CAN/LiRh$_2$O$_4$ amount is shown in Figure 1(a). 

\begin{figure}
    \includegraphics[width=0.9\textwidth]{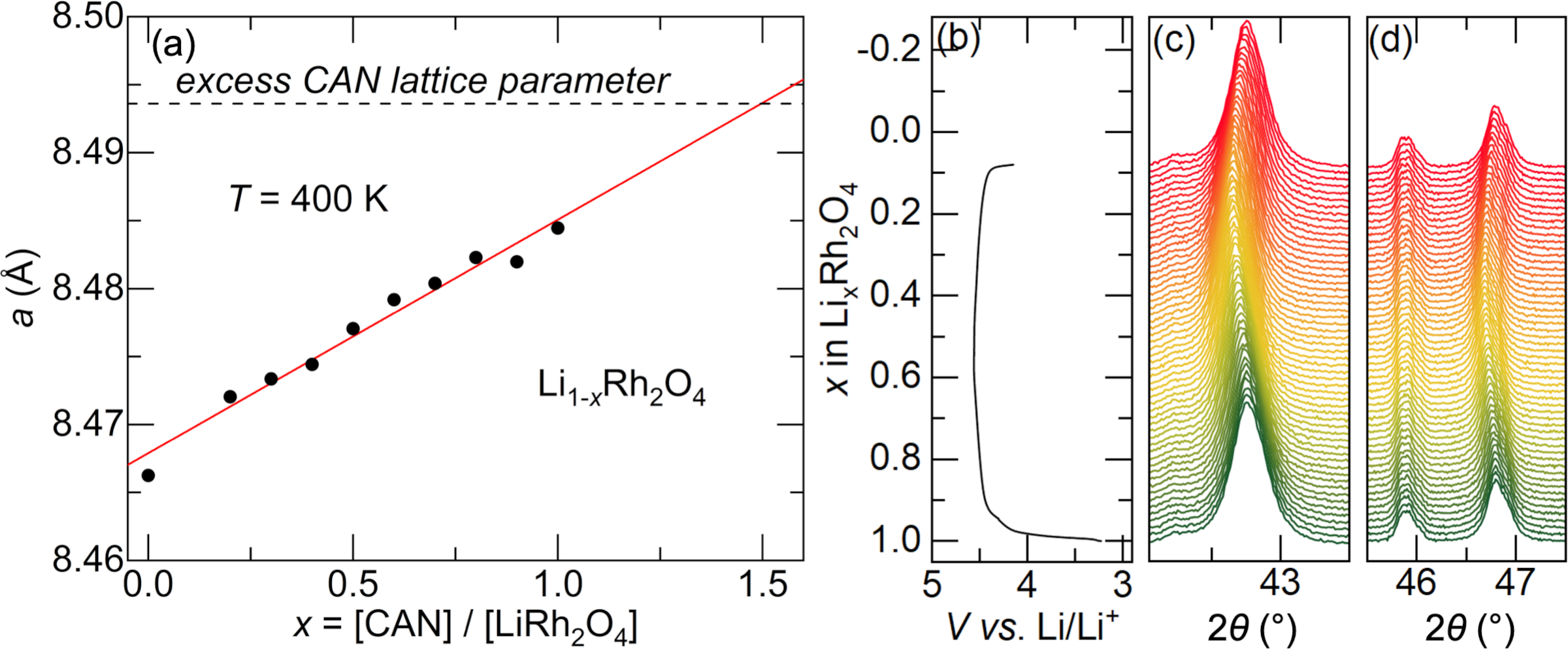}
    \caption{(a) Lattice parameter of cubic Li$_{1-x}$Rh$_2$O$_4$ as a function of \textit{x}, where \textit{x} is the relative molar amount of CAN to LiRh$_2$O$_4$ in each reaction. The trend is linear and intercepts with the $\lambda$-RhO$_2$ lattice parameter at \textit{x} = 1.5. (b)-(d) Results of operando X-ray diffraction measurements. The cell voltage and lattice parameter of Li$_{1-x}$Rh$_2$O$_4$ increases on charging (removal of Li$^+$ from the lattice) up to approximately \textit{x} = 0.45, beyond which the voltage and lattice parameter begins to decrease.}
\end{figure}

The lattice parameter of the cubic Li$_{1-x}$Rh$_2$O$_4$ cell increases approximately linearly with increasing targeted \textit{x}. However, $\lambda$-RhO$_2$, prepared with CAN in excess, has a cubic lattice parameter at \textit{T} = 400 K that is achieved when using a 1.5CAN:1.0LiRh$_2$O$_4$ molar ratio, suggesting that the above delithiation reaction is incomplete. Further work is required to understand the reaction mechanism, including the thermodynamic and kinetic barriers faced under these strong oxidizing conditions.

We also attempted to synthesize Li$_{1-x}$Rh$_2$O$_4$ phases with variable Li content using other reagents and electrochemistry. Reactions in either Br$_2$ or I$_2$ solutions in acetonitrile, common oxidative deintercalation agents \cite{Miyazaki1983,Neilson2012,Wizansky1989,Chamorro2018p}, did not yield any noticeable differences in diffraction patterns of Li$_{1-x}$Rh$_2$O$_4$ targeted samples \textit{vs.} LiRh$_2$O$_4$, nor significant variations in measurements of the low temperature physical properties. We also prepared an electrochemical cell of LiRh$_2$O$_4$ vs. Li-metal in order to test electrochemical delithiation. Cells were discharged to 1 V (lithiation) and then charged to 3 V (delithiation). As shown in Figure 1(b)$-$(d), the voltage of the cell quickly approaches an extended plateau at 4.5 V with a voltage profile similar to that observed in Mn- and Ni- containing spinels, albeit at a lower voltage \cite{CasasCabanas2016}. The removal of lithium from Li$_{1-x}$Rh$_2$O$_4$ results in an increase in the lattice parameter upon delithiation, marked by a shift in the diffraction peaks to lower angles. However, at approximately Li$_{0.55}$Rh$_2$O$_4$, both the voltage and lattice parameter begin to anomalously decrease, likely due to a breakdown of the cell electrolyte. As such, the electrochemical method employing an LiPF$_6$ electrolyte is insufficient in removing more than 45\% Li from the structure. Results from refinements of the operando X-ray diffraction measurements are shown in the supplementary information (Figure S1) and further indicate that $\lambda$-RhO$_2$ cannot be obtained electrochemically, as revealed by the absence of XRD reflections associated with the tetragonal $\lambda$-RhO$_2$ phase, discussed in more detail below.

\subsection{Average Structure, and Structure Evolution}

Rietveld refinements were performed on X-ray diffraction data sets collected on $\lambda$-RhO$_2$ obtained via chemical delithiation at the 11-BM beamline at Argonne National Laboratory between \textit{T} = 100 and 400 K. Data and fits for these two temperatures are shown in Figure 2, along with the cubic and tetragonal structures of $\lambda$-RhO$_2$.

\begin{figure}[h]
    \includegraphics[width=0.65\textwidth]{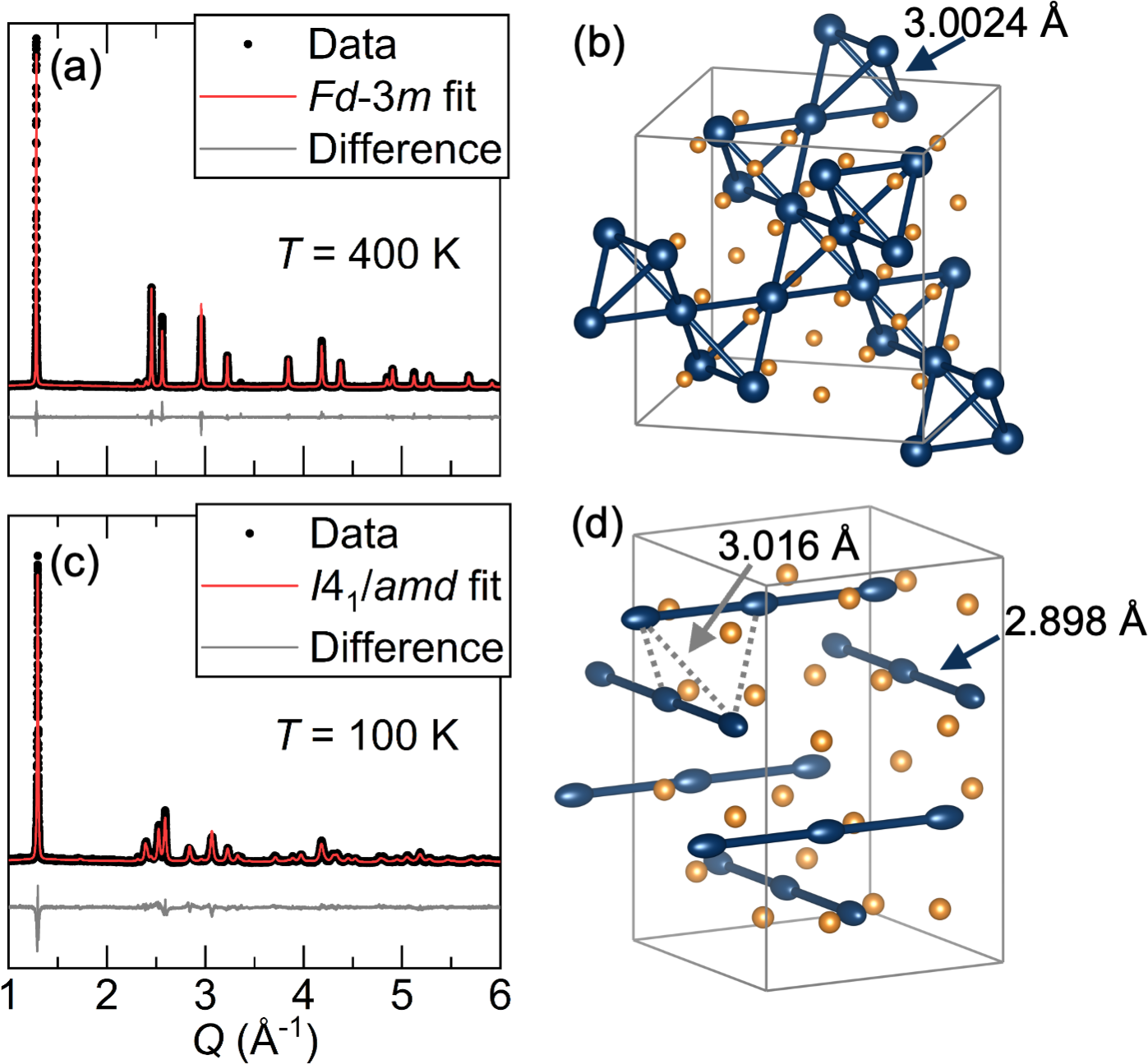}
    \caption{(a),(c) Synchrotron X-ray diffraction patterns collected at \textit{T} = 400 K and \textit{T} = 100 K, respectively, along with Rietveld refinement fits. (b) The structure of cubic $\lambda$-RhO$_2$, demonstrating the three-dimensional pyrochlore network of Rh cations. (d) The low-temperature, \textit{I}4$_1$/\textit{amd} structure, where only the shortest Rh-Rh distances are shown. Rh cations are shown as displacement ellipsoids to highlight the anisotropy of the refined displacement parameters.}
\end{figure}

At \textit{T} = 400 K, $\lambda$-RhO$_2$ forms in the prototypical cubic \textit{Fd}$\bar{3}$\textit{m} spinel structure. Refinements of anisotropic displacement parameters for rhodium and oxygen do not result in a significant increase of the goodness of fit, implying that in the cubic phase, displacement parameters are likely isotropic. A structural phase transition occurs between \textit{T} =\,320\,$-$\,340 K which reduces the average structure from cubic to tetragonal \textit{I}4$_1$/\textit{amd}. Fit parameters can be found in the supplementary information (Tables S1 and S2).

A consequence of the cubic-to-tetragonal structural phase transition is the formation of rhodium \textit{xy}-chains along either \textit{a} or \textit{b}. In the cubic phase, the nearest-neighbor Rh$-$Rh distance is 3.002(1) $\mathrm{\AA}$, whereas in the tetragonal phase, Rh$-$Rh intrachain distances become 2.898(2) $\mathrm{\AA}$ and Rh$-$Rh interchain distances become 3.016(3) $\mathrm{\AA}$. This distortion is likely due to a Jahn-Teller distortion of the low-energy Rh$^{4+}$ $t_{2g}$ orbital manifold, where the orbital degeneracy is lifted through a lowering of the $d_{xz}$ and $d_{yz}$ orbitals relative to $d_{xy}$. Refinements of the rhodium anisotropic displacement parameters indicate a predilection for Rh displacements within the \textit{xy}-chains, with a maximal displacement \textit{B}$_{22}$ = 0.862(1) $\mathrm{\AA}$, as opposed to \textit{B}$_{11}$ = 0.287(2) $\mathrm{\AA}$. Figure 2(d) shows the low temperature tetragonal structure of $\lambda$-RhO$_2$, where the \textit{xy}-chains have been highlighted as well as the anisotropic displacement parameters that are larger within the chain axes \textit{vs.} any other direction.

\begin{figure}[h]
    \includegraphics[width=0.45\textwidth]{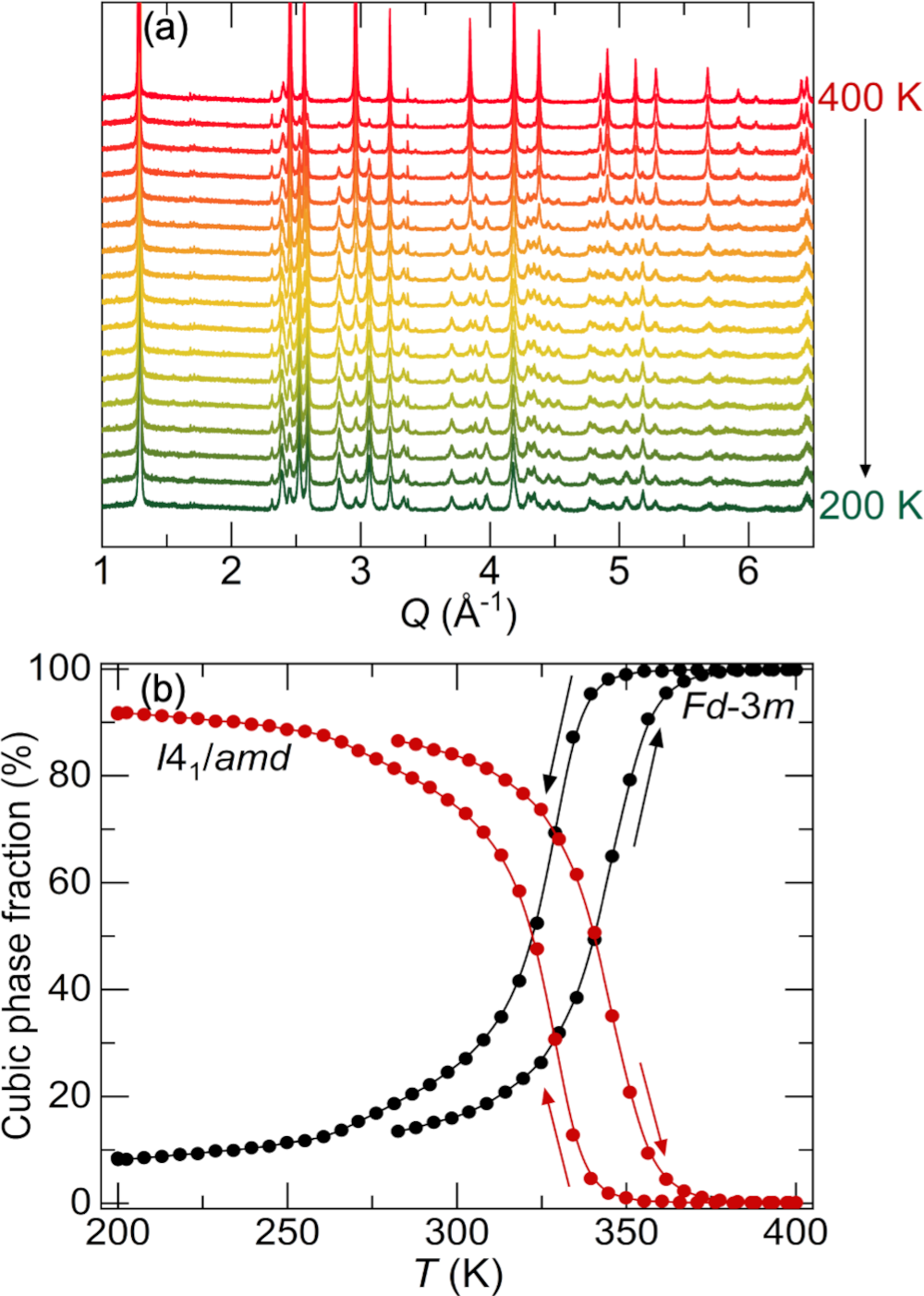}
    \caption{(a) Synchrotron X-ray diffraction patterns collected at various temperatures down to \textit{T} = 100 K. (b) The phase fractions of both cubic and tetragonal structures of $\lambda$-RhO$_2$. Approximately 8.4\% of the cubic phase remains at low temperature.}
\end{figure}

The structural phase transition is hysteretic in temperature, as demonstrated in Figure 3. Based on the first derivative of the phase fractions as a function of temperature, the transition is centered around $T_W$ = 345 K on warming and $T_C$ = 329 K on cooling. The \textit{c}$_{cubic}$/\textit{a}$_{cubic}$ = \textit{c}$_{tetragonal}$/$\sqrt{2}$\textit{a}$_{tetragonal}$ ratio at \textit{T} = 100 K, 1.08, indicates an 8\% departure from cubic symmetry across the transition. All samples show a remnant cubic phase at the lowest measured temperatures, on the order of 8-10\%, regardless of synthetic conditions. As demonstrated later with complementary solid-state NMR results, this remnant cubic phase could be related to a fraction of the sample that is not delithiated.

Long-range structural phase transitions have been observed in other spinel systems with electronic degrees of freedom on cations on the \textit{B}-site, such as CuIr$_2$S$_4$ \cite{Furubayashi1994,Hagino1994} and MgTi$_2$O$_4$ \cite{Isobe2002,Schmidt2004,Leoni2008}. In these compounds, which possess active spin, orbital, and charge degrees of freedom, structural transitions are observed due to the formation of molecular, non-magnetic units at low temperature, such as Ir$^{3+}$$-$ Ir$^{4+}$ octamers in the former \cite{Radaelli2002,Ishibashi2001} and helical Ti$^{3+}$$-$ Ti$^{3+}$ dimers in the latter \cite{Schmidt2004,Leoni2008}. In both of these compounds, a single phase transition occurs as in $\lambda$-RhO$_2$ (hysteretic in the former and non-hysteretic in the latter), whereas two phase transitions are observed in LiRh$_2$O$_4$. It is also instructive to compare these findings to those in the spinel magnetite Fe$_3$O$_4$, where a single transition (Verwey transition) is observed near \textit{T} = 120 K that is also the result from a complex coupling of Fe electronic degrees of freedom \cite{VERWEY1939,Anderson1956,Iizumi1982}. At the local scale, strong short-range correlations have been observed in both CuIr$_2$S$_4$ \cite{Bozin2019} and MgTi$_2$O$_4$ \cite{Torigoe2018} that preceed the long-range structural phase transition that are suggestive of dynamic short range fluctuations arising from electronic correlations. These have also been suggested in LiRh$_2$O$_4$ at temperatures \textit{T}$_{CO} <$ \textit{T}, and we discuss these in comparison to $\lambda$-RhO$_2$ in more detail in the following local structure section.

\begin{figure}[h]
    \includegraphics[width=0.65\textwidth]{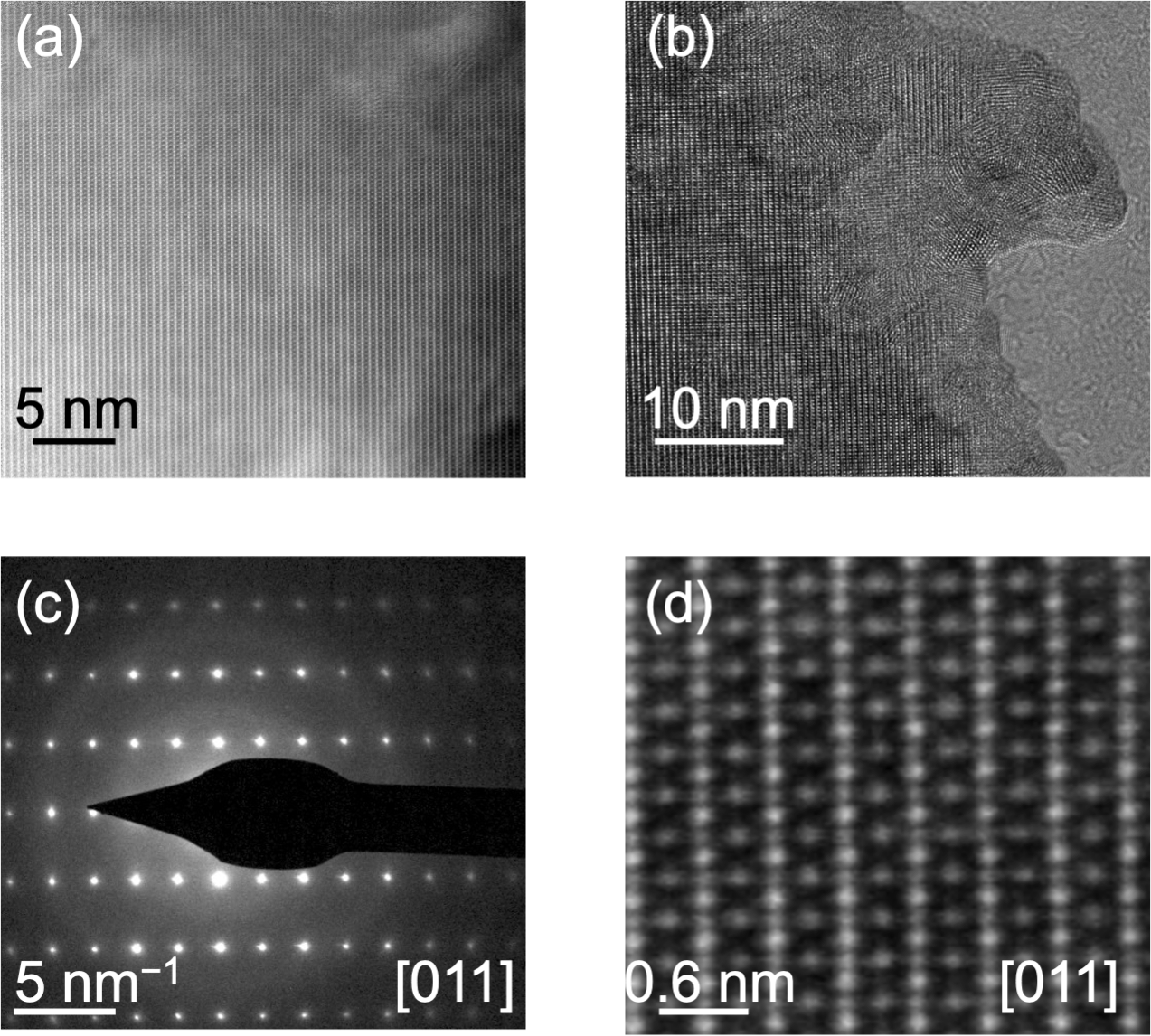}
    \caption{Electron microscopy characterization of $\lambda$-RhO$_2$ at room temperature. (a) HAADF-STEM image of of a single crystallite of $\lambda$-RhO$_2$, demonstrating sample homogeneity. (b) High-resolution TEM image of the single crystallite near the edge. Inhomogeneities can be observed near the edges of crystallites, possibly due to both cubic and tetragonal substructures owing to disparate Li content. (c) Electron diffraction pattern showing the tetragonal phase along [011]. (d) HAADF image of a single crystallite along the [011] zone axis.}
\end{figure}

In order to further study the average structure of $\lambda$-RhO$_2$, we employed transmission electron microscopy (TEM) and high-angle annular dark-field (HAADF) imaging measurements on polycrystalline samples, the results of which are shown in Figure 4. Position averaged convergent beam electron diffraction (PACBED) was performed to accurately determine the zone axis for high-resolution imaging. The bulk of each crystallite was found to be homogeneous and give rise to a well-defined diffraction pattern (Figures 4(a), (c)), with well defined and identifiable crystal planes (Figure 4(d)). The diffraction patterns lack any noticeable features beyond the main Bragg reflections.

Near the edges of the crystallites, non-uniformities are observed that depart from the homogeneous bulk. These homgeneities could be due to either an inhomogeneous distribution of Li throughout the particles, or a redistribution of Rh near the edges. The presence of lithium ultimately determines whether the structure is expected to be cubic or tetragonal, especially at room temperature, and the inhomogeneity observed by electron diffraction could arise from an admixture of cubic and tetragonal Li$_{1-x}$Rh$_2$O$_4$ phases near the edges of the crystallites, as in Figure 4(b). This hypothesis is supported by NMR observations discussed below.

\subsection{Local Structure Measurements}

Total scattering measurements were performed on $\lambda$-RhO$_2$ at the 11-ID-B beamline at Argonne National Laboratory between temperatures of \textit{T} = 100 and 400 K through the use of a liquid nitrogen cryostream. Analyses of this data up to high-\textit{Q}, in this case \textit{Q}$_{max}$ = 18 $\mathrm{\AA}$, allows for the extraction of the pair distribution function (PDF), which can provide information about atom-atom correlations via a Fourier transform of total scattering data, shown in Figure\,5. It offers a window into the local structure in materials and permits the study of distortions and structural transformations at the local bonding scale.

\begin{figure}
    \includegraphics[width=0.75\textwidth]{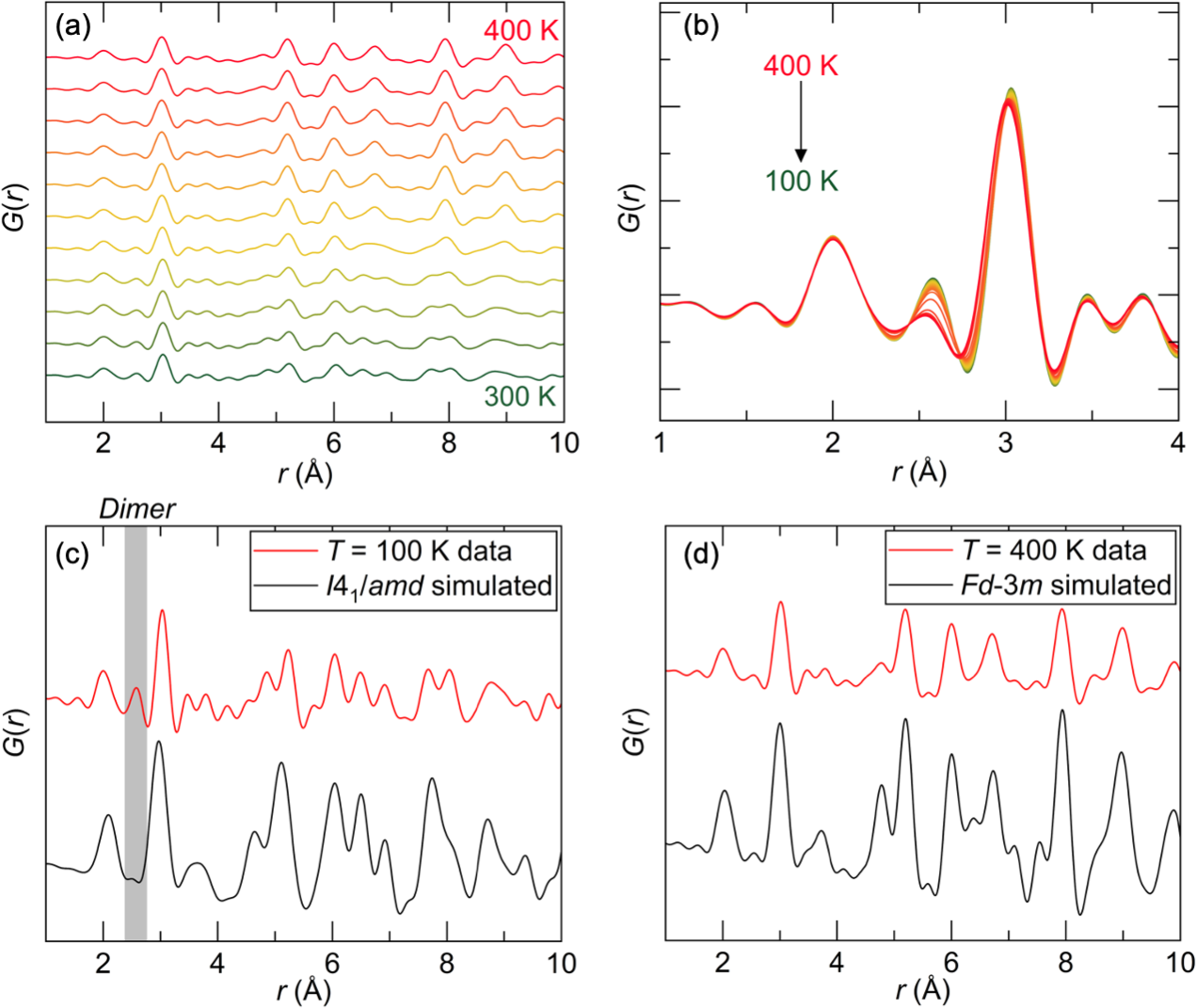}
    \caption{(a) Temperature-dependent X-ray pair distribution function measurements, demonstrating a dramatic change in the local structure of $\lambda$-RhO$_2$ across the $T_W$ = 345 K on warming and $T_C$ = 329 K structural phase transition. (b) Measured PDF patterns for $\lambda$-RhO$_2$, demonstrating the dimerization peak around 2.64 $\mathrm{\AA}$ that arises on cooling. (c)$-$(d) Measured \textit{vs.} simulated PDF patterns for the low and high temperature phases of $\lambda$-RhO$_2$. While the \textit{Fd}$\bar{3}$\textit{m} fit appears to match the data reasonably well at \textit{T} = 400 K, the \textit{I}4$_1$/\textit{amd} fit that captures the average structure does not match the local structure well.}
\end{figure}

As can be observed in Figure 5(a), the PDF patterns of $\lambda$-RhO$_2$ change intensely across the structural phase transition above room temperature at nearly all \textit{r} length scales. This is in agreement with a long-range structural phase transition, as the new low temperature cell is expected to possess vastly different atom-atom correlations compared to the high temperature cubic structure. However, as can be readily observed in Figure 5(c), unlike the cubic fit in Figure 5(d), the tetragonal \textit{I}4$_1$/\textit{amd} cell that reasonably fits the diffraction data cannot properly fit the PDF data. This suggests that the local structure of $\lambda$-RhO$_2$ differs from the average structure and suggests the presence of short-range correlations with a limited correlation length.

One pronounced difference between the observed low-\textit{T} PDF pattern (collected at \textit{T} = 100 K) and the simulated tetragonal PDF pattern (Figure 5(c)) is the presence of a peak that is centered around 2.64 $\mathrm{\AA}$, as demonstrated in Figure 5(b). A similar peak has been observed in the parent compound LiRh$_2$O$_4$ in the 2.70$-$2.75 $\mathrm{\AA}$ range \cite{Knox2013,Shiomi2022}, as well as in CuIr$_2$S$_4$ \cite{Radaelli2002,Bozin2011}, and has been attributed in both cases to short-range Rh$-$Rh dimerization (Ir$-$Ir in CuIr$_2$S$_4$). In LiRh$_2$O$_4$, in particular, this peak emerges in the PDF patterns even above \textit{T}$_{JT}$ (where LiRh$_2$O$_4$ is cubic), implying that dimerization is favored at the local scale regardless of the average structure. Dimerization in LiRh$_2$O$_4$ persists on cooling through both the orbital ordering (Jahn-Teller distortion) and charge ordering transitions, though it reaches a constant maximum value below the latter, implying a likely coupling of the spins to the charge degrees of freedom.

The dimerization peak observed in $\lambda$-RhO$_2$, which can be observed clearly in Figure 5 (b), differs from that in LiRh$_2$O$_4$ in that it does not appear above the long-range phase transition observed in average structure and physical properties measurements, indicating that dimers are not present in the high-temperature cubic structure, as shown in Figure 5(d). This implies that these dimers are primarily the result of the formation of Rh-Rh metal-metal bonds at low temperature, assisted by a Jahn-Teller distortion of the RhO$_6$ octahedra on cooling. Since $\lambda$-RhO$_2$ possesses mostly Rh$^{4+}$ (Rh$^{3.95+}$ assuming Li$_{0.1}$Rh$_2$O$_4$), there are no charge degrees of freedom that could give rise to a long-range charge ordered phase such as in LiRh$_2$O$_4$. 

Given that these dimers appear only in the local structure and not in the average, it is possible they are either dynamically fluctuating or static and disordered. The observation of large anisotropic displacement parameters for Rh in the synchrotron X-ray diffraction data, however, suggests that the dimers are only present at the local scale. In our refinements of the global structure, they manifest as large displacement parameters along the \textit{xy}-chains, and form locally along these chains on cooling. Given that the chains are made of Rh$^{4+}$ species with an effective spin \textit{S}$_{\mathrm{eff}}$ = 1/2 , they are likely susceptible to an orbitally driven Peierls instability.

$^{7}$Li solid-state NMR measurements were performed on $\lambda$-RhO$_2$ and LiRh$_2$O$_4$ in order to further probe the local structure around Li$^+$, as well as to quantify any remnant amount of Li$^+$ cations within $\lambda$-RhO$_2$ after chemical delithiation . The room temperature $^{7}$Li NMR spectrum collected on LiRh$_2$O$_4$ exhibits two resonances at 0 ppm (\textit{T}$_2$-weighted integral 3\%) and 50 ppm (97\%), as shown in Figure 6(a). The 0 ppm peak is assigned to diamagnetic impurities (\textit{e.g.}, Li$_2$CO$_3$, LiOH, LiNO$_3$, and Li$_2$O) at the surface of the LiRh$_2$O$_4$ particles. The 50 ppm peak likely corresponds to Li in the LiRh$_2$O$_4$ structure, which at room temperature occupies solely the tetrahedrally-coordinated \textit{A}-site, and therefore gives rise to a single resonance. This relatively large shift may arise from the Knight shift interaction, as well as the Fermi contact interaction, between the unpaired electrons on Rh$^{3+/4+}$ cations and the vacant Li$^+$ \textit{s}-orbitals\cite{Grey2004}.

\begin{figure}
    \includegraphics[width=0.85\textwidth]{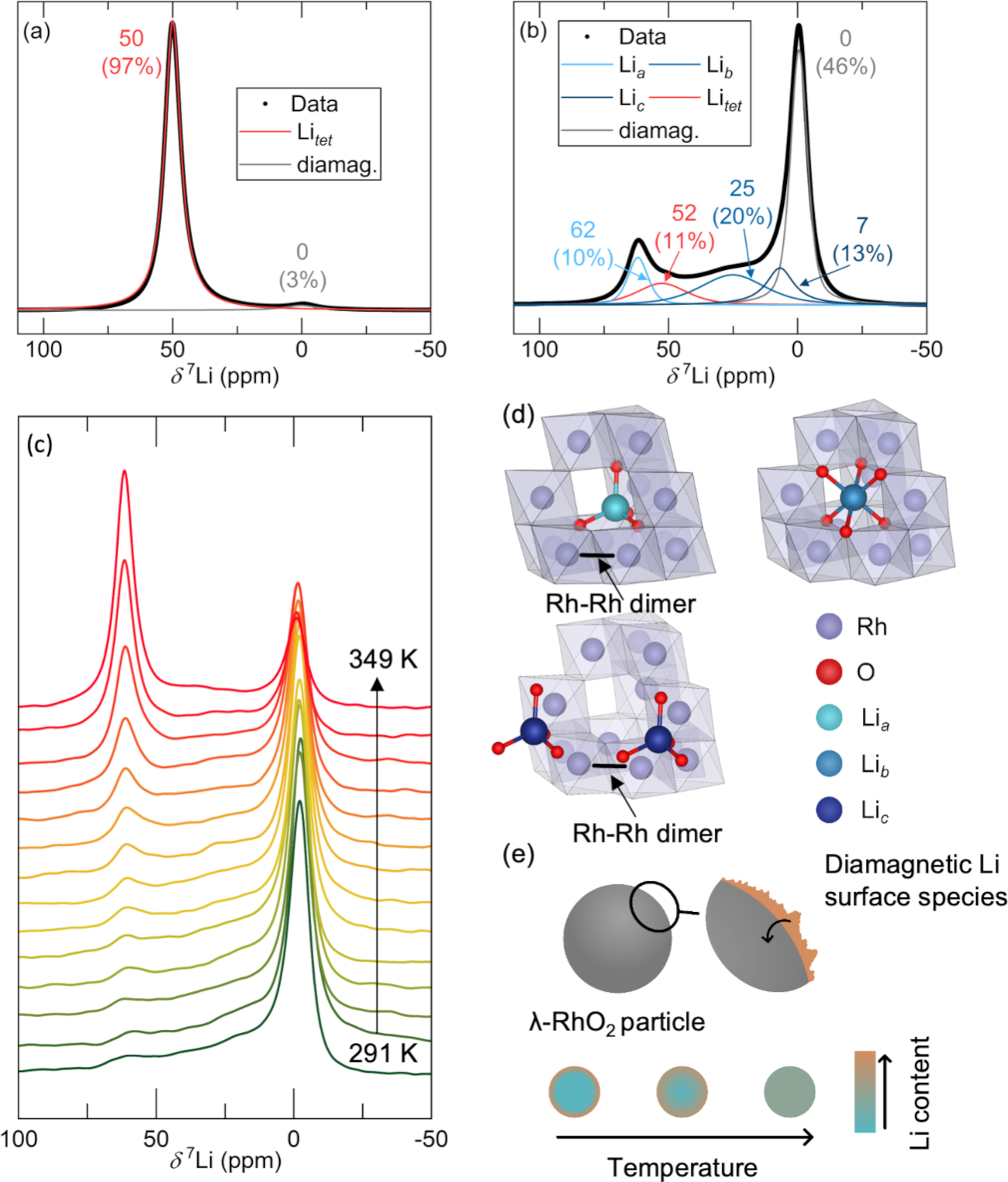}
    \caption{$^7$Li NMR of (a) LiRh$_2$O$_4$ and (b) $\lambda$-RhO$_2$, recorded at 2.35 T under a MAS speed of 60 kHz, corresponding to a sample temperature of 318 K. (c) Variable-temperature $^7$Li NMR of $\lambda$-RhO$_2$ (2.35 T field, MAS speed 60 kHz) between 291 and 349 K. (d) The three local Li environments in $\lambda$-RhO$_2$ are indicated, with the Rh$-$Rh dimer position indicated for the two tetrahedral interstitial sites Li$_a$ and Li$_c$, as well as the octahedral interstitial site Li$_b$. (e) Illustration of the migration of Li from the surface-based diamagnetic species into bulk $\lambda$-RhO$_2$, with schematic lithiation gradients below, at and above the transition temperature.}
\end{figure}

Delithiation of LiRh$_2$O$_4$ to $\lambda$-RhO$_2$ results in a complex $^{7}$Li NMR spectrum comprising at least five overlapping resonances at roughly 0 (46\%), 7 (13\%), 25 (20\%), 52 (11\%), and 62 ppm (10\%), as shown in Figure 6(b). The peak at 0 ppm is again assigned to diamagnetic species, and the higher intensity of this resonance in the data obtained on $\lambda$-RhO$_2$ as compared to LiRh$_2$O$_4$ is expected, as the synthesis likely results in the nucleation of Li salts at the surface of the $\lambda$-RhO$_2$ particles during chemical delithiation. The peak at \textit{ca.} 50 ppm is assigned to Li in an LiRh$_2$O$_4$-like (i.e. a pristine-like) environment. Integration of this 50 ppm resonance suggests that about 10$-$11\% of the Li in the $\lambda$-RhO$_2$ sample occupies tetrahedral \textit{A}-sites in LiRh$_2$O$_4$-like domains.

The three remaining signals at 7, 25, and 62 ppm have approximate integral ratios of 1:2:1, suggesting that the resonances at 7 ppm and 62 ppm correspond to the tetrahedral Li interstitial sites (Wyckoff site 4\textit{a}), and the shift at 25 ppm corresponds to octahedral Li interstitials (Wyckoff site 8\textit{c}) in $\lambda$-RhO$_2$, respectively. These are shown in Figure 6(d), where Li$_a$ and Li$_c$ correspond to the tetrahedral Li cations and Li$_b$ corresponds to the octahedral Li. We propose that the origin of the two unique tetrahedral resonances Li$_a$ and Li$_c$ stems from the proximity of these sites to the Rh-Rh dimers that form along the \textit{xy}-chains. By analogy with the shifts of Li species near 
Ru-Ru dimers in Li$_2$RuO$_3$ \cite{Reeves2019, Kimber2014, Park2016}, it is expected that Li occupying a tetrahedral site between dimers will experience a smaller shift than Li occupying a site at the edges of the dimers, as the Rh orbitals involved in Rh-Rh metal-metal bond formation are partially spin-quenched, resulting in a smaller effective electron spin moment for the Li nucleus to interact with, while those pointing away from the dimer will contain a larger effective spin moment. The absolute integrals of the $^7$Li NMR spectra (sample mass-normalized and \textit{T}$_2$-weighted) obtained on the LiRh$_2$O$_4$ and $\lambda$-RhO$_2$ samples suggest that the chemically-delithiated sample contains approximately 16(1)\% of the Li in the pristine sample.

Variable-temperature NMR measurements were also performed on LiRh$_2$O$_4$ (Figure S2) and $\lambda$-RhO$_2$ samples (Figure 6(c)). These measurements were carried out at sample temperatures between \textit{T} = 291 and 349 K. In LiRh$_2$O$_4$, the 50 ppm shift remains temperature-independent, suggesting metallic behavior and a shift that is dominated by the Knight shift interaction (Figure S2(a)). For the $\lambda$-RhO$_2$ sample, the extensive overlap between the aforementioned Li$_a$, Li$_b$, Li$_c$, and Li in LiRh$_2$O$_4$ resonances prevents us from tracking subtle chemical shift variations with temperature, but to a first approximation, those shifts do not exhibit a significant temperature dependence. Interestingly, while the overall $^7$Li NMR signal intensity obtained from the $\lambda$-RhO$_2$ sample remains roughly constant with temperature, the relative intensities of the Li$_a$, Li$_b$, Li$_c$, Li in LiRh$_2$O$_4$, and diamagnetic resonances vary drastically between 349 K and 327 K (Figure S2(e)). Over this temperature range, the intensity of the Li$_a$ resonance increases at the expense of the diamagnetic signal as temperature increases. This suggests Li chemical exchange or a transfer of Li population from the diamagnetic Li-containing salts accumulated at the surface of the $\lambda$-RhO$_2$ particles to the bulk, and in particular to the tetrahedrally-coordinated Li$_a$ environments of the spinel structure. The onset of this redistribution of Li populations appears to occur between 322 and 331 K, which is concomitant with the phase transformation from \textit{I}4$_1$/\textit{amd} to \textit{Fd}$\bar{3}$\textit{m}. Hence, we speculate that the large quantity of Li in diamagnetic surface species (presumably generated during synthesis) is driven into bulk Rh$_2$O$_4$ on heating, resulting in an increased Li content in the outer layers of the $\lambda$-RhO$_2$ particles near room temperatures (Figure 6(e)). We also observe a drop in intensity of all signals apart from the diamagnetic Li from 322 to 327 K, which we tentatively ascribe to a shortened $T_2$, analogous to the loss in signal seen in LiCoO$_2$ near the metal-to-insulator transition temperature\cite{menetrier_insulator-metal_1999}. Whilst the precise origin of this enhanced transverse dephasing of nuclear magnetization is unclear, it appears connected to this transition; this poses an interesting avenue for future research. We suggest that the origin of the phase transformation from tetragonal \textit{I}4$_1$/\textit{amd} to cubic \textit{Fd}$\bar{3}$\textit{m} is mediated by the changing Li composition over this temperature range. Those results are consistent with our observations in TEM measurements, where the bulk of the crystallites appeared homogeneous and with minimal disorder, as opposed to the crystallite edges which appeared greatly disordered and likely containing admixtures of both cubic Li$_{1-x}$Rh$_2$O$_4$ and tetragonal $\lambda$-RhO$_2$.

In summary, the presence of short Rh-Rh bonds is evidenced in both PDF and NMR, despite the predicted average \textit{I}4$_1$/\textit{amd} structure for $\lambda$-RhO$_2$ that does not accommodate these bonds. Given this disparity between the average and local structures, these dimers are either static and disordered, or dynamically fluctuating on a timescale comparable to the diffraction measurement. In LiRh$_2$O$_4$, this has been suggested in the temperature range of $T_{CO} < T < T_{JT}$, where a dimerization peak appears in the PDF despite an average tetragonal \textit{I}4$_1$/\textit{amd} structure \cite{Knox2013,Shiomi2022}. Alternatively, the presence of a small amount of remaining Li post-delithiation within the $\lambda$-RhO$_2$ structure could prevent a long-range phase transition to an ordered state due to the disorder induced by this Li, thus Rh-Rh bonding is only observed in local probes.

\subsection{Physical Properties Measurements.}

\begin{figure}
    \includegraphics[width=0.8\textwidth]{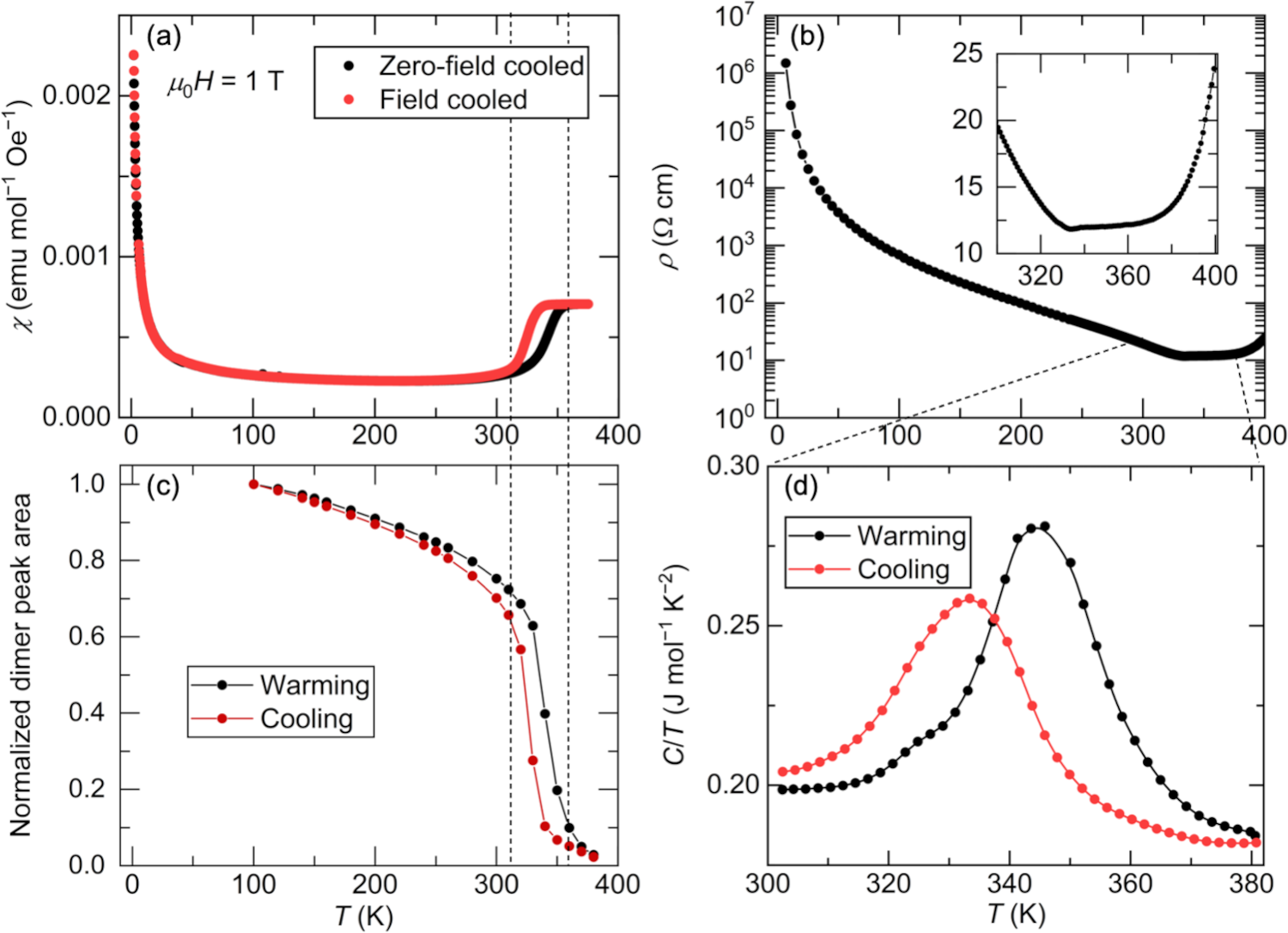}
    \caption{Temperature-dependent physical properties measurements of $\lambda$-RhO$_2$, as well as the normalized dimer peak area from PDF measurements. (a) dc magnetic susceptibility collected under an applied field of $\mu_0H$ = 1 T. (b) Resistivity measurement under zero applied field on cooling and warming. (c) Normalized dimer peak area from PDF measurements, demonstrating it to be simultaneous with the metal-to-insulator transition and transition to a non-magnetic state. (d) Heat capacity measurement under zero applied field on cooling and warming, demonstrating the transition.}
\end{figure}

Physical properties measurements were carried out on $\lambda$-RhO$_2$ and LiRh$_2$O$_4$, including temperature-dependent magnetic susceptibility, electrical resistivity, and heat capacity, as shown in Figure 7.

The dc magnetic susceptibility of $\lambda$-RhO$_2$, shown in Figure 7(a), shows a drop at \textit{T}$_{M-W}$ = 342 K on warming and \textit{T}$_{M-C}$ = 325 K on cooling, consistent with the transition in diffraction data. This transition is indicative of a transition into a non-magnetic state, such as in LiRh$_2$O$_4$ below the charge ordering transition. In contrast to LiRh$_2$O$_4$, however, $\lambda$-RhO$_2$ shows only one, hysteretic transition. The formation of a non-magnetic state is consistent with the formation of Rh-Rh dimers, as the individual \textit{S} = 1/2 Rh$^{4+}$ cations pair up to form \textit{S}$_{\mathrm{eff}}$ = 0 singlets. Curie-Weiss behavior was not identified at any temperature region, including above \textit{T} = 400 K (Figure S3 in the supplementary information), suggesting the absence of localized moments.  

Measurements of electrical resistivity, shown in Figure 7(b), demonstrate a metal-to-insulator transition on cooling across the transition. The resistivity increases by five orders of magnitude below the transition, but cannot be fit to either an exponentially activated or variable range hopping model. Given the fact that $\lambda$-RhO$_2$ is a metastable compound, annealing or sintering pellets for measurements at high temperatures is not a possibility. As such, measurements were performed solely on cold-pressed pellets, which likely affects the measurements through poor conductivity across grain boundaries. Nevertheless, as can be seen in the inset, the resistivity increases with increasing temperature above the transition, suggesting a metallic state.

A metal-to-insulator transition occurs in LiRh$_2$O$_4$\cite{Okamoto2008}, as well as in the spinels CuIr$_2$S$_4$ \cite{Furubayashi1994,Hagino1994,Nagata1994}, MgTi$_2$O$_4$ \cite{Isobe2002,Zhu2014}, and LiV$_2$O$_4$ (under pressure and doped)\cite{Browne2020,Kawakami1986,Onoda1997}. A mechanism that has been proposed to explain the transition in such compounds is the orbitally-driven Peierls state, whereby strong anisotropic orbital interactions between adjacent sites establish quasi-one-dimensional chains within the pyrochlore lattice\cite{Khomskii2005}. These chains are then individually susceptible to a Peierls instability, and the pyrochlore network distorts to form either complex molecular clusters, such as the octamers in CuIr$_2$S$_4$ \cite{Radaelli2002}, or dimers, such as in MgTi$_2$O$_4$ \cite{Schmidt2004,Yang2008}. This mechanism likely also explains the observed transition in $\lambda$-RhO$_2$, though we do not observe a long-range crystal structure with dimers for this compound. Figure 7(c) shows the normalized dimer peak area from PDF measurements, demonstrating that the onset of the metal-to-insulator and the magnetic-to-non-magnetic transitions are concomitant with the formation of short-range dimers. It appears as though dimerization is strongly favored at the local scale through local bonding interactions. However, we do not rule out a lower symmetry average structure that could accommodate either dimers or some larger \textit{n}-unit non-magnetic cluster arrangement, such as the octamer ground state in CuIr$_2$S$_4$ \cite{Radaelli2002} or trimers in RuP \cite{Hirai2022}. 

In theory, one would expect $\lambda$-RhO$_2$ and LiRh$_2$O$_4$ to display behavior closer to CuIr$_2$S$_4$ than to MgTi$_2$O$_4$ and LiV$_2$O$_4$, owing to the general similarity of Rh/Ir chemistry and the comparatively strong spin-orbit coupling in Rh vs. Ti/V. However, in both LiRh$_2$O$_4$ and $\lambda$-RhO$_2$, spin-orbit coupling appears to play a smaller role in relation to the Jahn-Teller instability of the low-energy Rh$^{4+}$ \textit{t}$_{2g}$ orbital manifold. As such, a cubic-to-tetragonal phase transition occurs in both that establishes a ground state structure with Rh-Rh chains along the \textit{a}- or \textit{b}-crystallographic directions (\textit{xy}-chains). Dimerization can then naturally occur along these chains composed of half-filled Rh$^{4+}$ via an orbitally-driven Peierls state \cite{Khomskii2005,Bozin2019, Britto2015,Hiroi2015}. In the following section, we examine the impact of correlations and spin-orbit coupling on the electronic structure via calculations based on density functional theory.

\subsection{Electronic Structure}

Calculations based on density functional theory (DFT) were performed on LiRh$_2$O$_4$ and $\lambda$-RhO$_2$ using experimentally-derived structural parameters. $\lambda$-RhO$_2$ was treated as having no lithium on the \textit{A}-site. Calculations were done on both the cubic \textit{Fd}$\bar{3}$\textit{m} and tetragonal \textit{I}4$_1$/\textit{amd} cells with and without structural relaxation using the the Perdew-Burke-Ernzerhof (PBE) exchange functional\cite{Perdew2008}.  

Based on PBE-relaxed structures, the energy per formula unit of rutile RhO$_2$ is more than 0.5 eV lower than that of both cubic and tetragonal $\lambda$-RhO$_2$. The greater computed thermodynamic stability of the rutile phase is consistent with $\lambda$-RhO$_2$ being a kinetically trapped phase accessible only through low-temperature, oxidative delithiation. This is further corroborated by the observation that heating $\lambda$-RhO$_2$ in air above \textit{T} = 500 K results in decomposition to rutile RhO$_2$ and Rh$_2$O$_3$.

\begin{figure}[h]
    \includegraphics[width=0.8\textwidth]{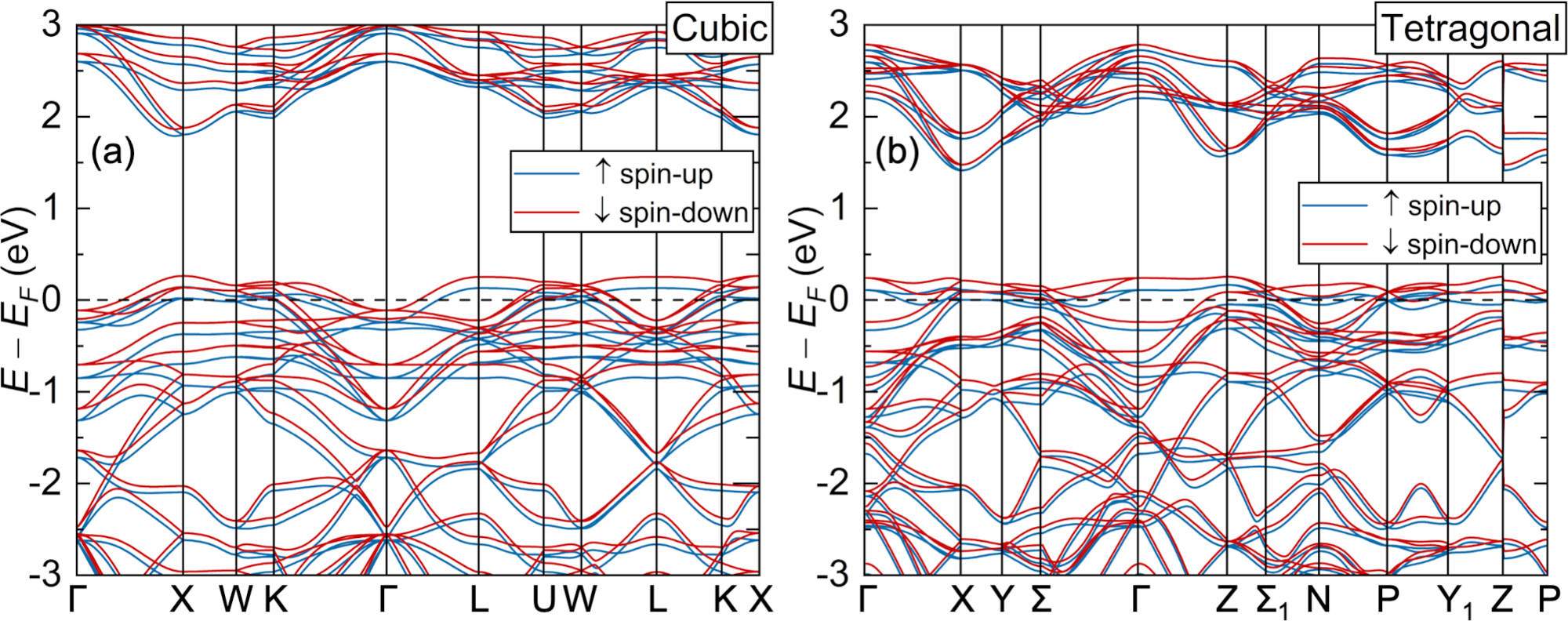}
    \caption{(a) Electronic band structure of $\lambda$-RhO$_2$, calculated using PBE and $U_{\mathrm{eff}}$ = 4 eV. Orange and blue bands correspond to spin-up and spin-down states, respectively. (b) Density of states plot, demonstrating a gap at the Fermi level along with strong spin-polarization, likely due to the absence of dimers in the calculation.}
\end{figure}

Structural relaxations of the cubic and tetragonal phases of $\lambda$-RhO$_2$ were performed with and without Hubbard $U_{\mathrm{eff}}$ values applied to the Rh 4$d$ bands and/or spin-orbit coupling (SOC). The results, shown in the supplementary information (Table S3), reveal that the tetragonal structure is only stabilized relative to the cubic one upon inclusion of a $U_{\mathrm{eff}} >$ 1 eV term. However, the inclusion of an SOC term results in a destabilization of the tetragonal phase. These results suggest that the tetragonal phase is enabled by correlations that drive the formation of the \textit{xy}-chains, which exist on a higher energy scale compared to spin-orbit coupling. 

A gap in the calculated band structure for the low-temperature tetragonal phase can be observed for $U_{eff} \geq$ 3 eV, as demonstrated in the supplementary information (Figure S4). However, these calculations do not include any dimerization, which would naturally provide a gap-opening mechanism. As such, the predicted band structures in the absence of an applied, non-zero U, shown in Figure 8, predict metallicity and substantial spin-polarization. Applying a non-zero $U_{\mathrm{eff}}$ offers an avenue toward localization, though strong spin polarization is still observed. Our findings suggest that the spin polarization observed in our calculations stems from a natural instability toward either dimerization or magnetic order, and that the low temperature insulating state observed in experiment is likely the result of dimer-induced localization. Naturally, our calculations do not incorporate dimers, as these calculations have been performed on the long-range structures derived from fits to average structure measurements. Should a lower-symmetry, long-range structure exist that accomodates for the Rh-Rh dimers in $\lambda$-RhO$_2$, calculations of its band structure would likely relieve this apparent inconsistency between DFT and experiment.

We conclude our discussion of the structure and properties of $\lambda$-RhO$_2$ by considering it in comparison to other spinel compounds, especially CuIr$_2$S$_4$ and MgTi$_2$O$_4$, the only other spinels that have been found to undergo metal to non-magnetic insulator transitions without doping or external pressure, as well as the aforementioned pyrochlore iridates. Naively, one would expect both LiRh$_2$O$_4$ and $\lambda$-RhO$_2$ to display similar behavior as in CuIr$_2$S$_4$ rather than MgTi$_2$O$_4$. LiRh$_2$O$_4$, in particular, possesses Rh$^{3+/4+}$ just as CuIr$_2$S$_4$ has Ir$^{3+/4+}$. However, SOC plays a much larger role in establishing the single-ion physics of Ir compounds compared to Rh compounds, and as our results suggest, SOC indeed plays a near negligible role in establishing the ground state of $\lambda$-RhO$_2$. As such, these systems are closer to MgTi$_2$O$_4$ where SOC plays a minor role in relation to the single-ion orbital instability and gives rise to a long-range dimerized ground state. $\lambda$-RhO$_2$, therefore, presents an avenue toward the study of competing interactions in a 4\textit{d} transition metal with a rare oxidation state.

\section{Conclusion}

$\lambda$-RhO$_2$ represents a platform to study the interplay of orbital and spin degrees of freedom of 4\textit{d}$^5$ cations on a pyrochlore lattice. We have synthesized this new Rh$^{4+}$ oxide using ceric ammonium nitrate, a heavily understudied, powerful oxidizer with wide-ranging future applications in the low temperature, oxidative deintercalation of extended solids. Our measurements indicate the presence of short-range Rh-Rh dimers that arise from metal-metal bonding at the local scale that do not crystallize in the long-range average structure. These dimers arise across a hysteretic phase transition at $T_W$ = 345 K on warming and $T_C$ = 329 K on cooling, which is concurrent to a metal-to-insulator transition and a magnetic to non-magnetic transition. Our results inspire the search for other possible quantum materials in frustrated lattices made up of transition metals with uncommon, high oxidation states.

\section{Methods}

\textbf{Synthesis.} Polycrystalline samples of $\lambda$-RhO$_2$ were prepared using soft chemical techniques. First, LiRh$_2$O$_4$ spinel precursor was synthesized in evacuated silica tubes using stoichiometric amounts of Li$_2$O$_2$ and Rh$_2$O$_3$, as previously reported \cite{Okamoto2008}. Physically separated PbO$_2$, which decomposes at high temperatures via PbO$_2$ $\rightarrow$ PbO + $\frac{1}{2}$O$_2$, was used to generate high oxygen pressures within reaction tubes. Powders were characterized via in-house X-ray diffraction.

LiRh$_2$O$_4$ powders were then stirred in aqueous solutions of ceric ammonium nitrate for 48 hours, followed by vacuum filtration and drying in air. Stoichiometric amounts of ceric ammonium nitrate were used to target specific stoichiometries with formula Li$_{1-x}$Rh$_2$O$_4$. A ten-fold molar excess of ceric ammonium nitrate was used to synthesize the end-member, $\lambda$-RhO$_2$.

\textbf{X-ray structural characterization.} High resolution synchrotron powder X-ray diffraction (XRD) data was collected at the 11-BM beamline at the Advanced Photon Source (APS) at Argonne National Laboratory, using an incident wavelength of 0.4590443 $\mathrm{\AA}$. Data were collected at temperatures ranging from \textit{T} = 100 K to 400 K. Powder XRD data was also collected in-house using a Panalytical Empyrean diffractometer employing Cu K$\alpha$ X-rays in a Bragg-Brentano geometry.

Pair distribution function (PDF) datasets were collected at the 11-ID-B beamline at APS, using an incident wavelength of 0.2116 $\mathrm{\AA}$. Data were collected at temperatures ranging from \textit{T} = 100 K to 400 K, and PDF patterns were extracted with a maximum momentum transfer of $Q_{max} = 18$ $\mathrm{\AA}$. Modeling and fitting of the XRD and PDF data was performed using TOPAS Academic. 

\textbf{Physical properties measurements.} Magnetic susceptibility measurements were carried out on powder samples in a Quantum Design Magnetic Property Measurement System (MPMS3). Resistivity and heat capacity measurements were performed using a Quantum Design 14 T Dynacool Physical Property Measurement System (PPMS). Resistivity measurements were performed via the four probe method on cold pressed pellets of polycrystalline sample.

\textbf{Transmission Electron Microscopy.} TEM samples were made by first preparing a suspension through the mixing of $\lambda$-RhO$_2$ powder in water, and then drop casting the particle suspension on a TEM grid. The grid was subsequently dried in air below 80 $^{\circ}$C for approximately 2 minutes before being inserted in a Spectra 200 ThermoFisher Scientific transmission electron microscope equipped with an ultra-high-brightness gun source (X-CFEG) and six-fold astigmatism probe aberration corrector. The operating voltage is 200 kV, with a probe convergence semi-angle of 30 mrad. The detector semi-angle was set between 25 mrad to 200 mrad (camera length: 160 mm).

\textbf{$^7$Li solid state nuclear magnetic resonance.} Powder samples of LiRh$_2$O$_4$ and $\lambda$-RhO$_2$ were loaded into 1.3 mm diameter ZrO$_2$ magic angle spinning (MAS) rotors. $^7$Li NMR spectra were referenced to liquid LiCl in H$_2$O (1 M) at 0 ppm and acquired on a Bruker AVANCE (2.35 T) using a Bruker 1.3 mm MAS probe, a $\pi$/2 pulse length of 0.45 $\mu$s, and an MAS frequency of 60 kHz for “room temperature” spectra (i.e., no external heating or cooling applied) or 50 kHz for variable-temperature spectra. Rotor-synchronized Hahn-echo pulse sequences ($\pi$/2–$\tau$– $\pi$–$\tau$–acq.) were used to obtain spectra, the intensities of which were scaled by sample mass and number of scans. The recycle delay (2.5s; at least 5$T_1$) was set such that the bulk, paramagnetically shifted signal was recorded quantitatively and the diamagnetic signal due to surface-based impurities was suppressed; additional spectra with a longer recycle delay (25s) were recorded such that the diamagnetic signal was also recorded quantitatively. Sample temperatures were obtained from internal calibration of the $^{79}$Br shift of KBr \cite{Thurber2009}.

\textbf{Electrochemistry and operando X-ray diffraction. } Electrochemistry experiments were performed by casting electrodes made from a 80:10:10 (wt \%) ratio of LiRh$_2$O$_4$ : conductive carbon (TIMCAL Super P) : polyvinylidene fluoride (PVDF). The PVDF was first dissolved in N-methylpyrrolidone and mixed in a FlackTek speed mixer at 2000 rpm for 5 minutes. The conductive carbon and LiRh$_2$O$_4$ were ground in a mortar and pestle for 10 minutes and then added to the viscous mixture, forming a slurry. The slurry was mixed in the speed mixer for 10 minutes and later cast using a 200 $\mu$m blade. After 3 hours, the cast slurry was dried in a vacuum oven at 80 $^{\circ}$C overnight. The electrodes were punched into 10 mm diameter disks with loading between 2 and 3 mg cm$^{-2}$. The electrodes were brought into an Ar-filled glovebox (H$_2$O $<$ 0.1 ppm and O$_2$ $<$ 0.1 ppm) and assembled into Swagelok cells or Hohsen coin cells for electrochemical testing. A glass fiber separator (Whatman GF/D) was soaked in 1 M LiPF$_6$ in EC/DMC 50/50 v/v (Sigma-Aldrich) electrolyte, and a polished Li foil was used as the counter and reference electrode. Cells were discharged to 1 V and charged to 3 V using BioLogic potentiostats (VMP1 and VMP3). All measurements were carried out at room temperature.

Operando X-ray diffraction measurements were collected using a custom Swagelok-type cell with a Be window approximately 250 $\mu$m thick, allowing X-ray penetration into the cell while cycling. A pattern was collected every 20 minutes during cycling at a C/15 rate.

\textbf{Density functional theory.} First-principles electronic structure calculations were performed using the Vienna ab Initio Simulation Package (VASP) version 5.4.4. All calculations employed the Perdew-Burke-Ernzerhof (PBE) functional and projector-augemented wave potentials based on the v5.4 recommendations (Rh\_pv, O). The plane-wave energy cutoff was set to 520 eV and a 9 $\times$ 9 $\times$ 9 $\Gamma$-centered \emph{k}-point mesh was used to avoid incompatibility problems associated with the primitive cells of body-centered tetragonal structures. Electronic band structure and density of states calculations were performed on both the experimentally-derived structure (unrelaxed) and geometrically-optimized structures (relaxed) in which forces were converged within 10\textsuperscript{$-$5} eV/\AA. A \emph{k}-point path for the band structure was generated using the AFLOW online tool. All calculations had an energy convergence better than 10\textsuperscript{$-$6} eV.

\begin{acknowledgement} 
J.R.C. acknowledges support through the NSF MPS-Ascend Postdoctoral Fellowship (DMR-2137580). J.R.C, S.S., and S.D.W acknowledge support by the U.S. Department of Energy (DOE), Office of Basic Energy Sciences, Division of Materials Sciences and Engineering under Grant No. DE-SC0020305. E.N.B. and R.J.C. acknowledge and are grateful to the Spectroscopy Facility at UC Santa Barbara. S.S. and G.Z. acknowledge support by the U.S. Department of Energy under Grant No. DEFG02-02ER45994. The research reported here made use of the shared facilities of the Materials Research Science and Engineering Center (MRSEC) at UC Santa Barbara: NSF DMR-2308708. The UC Santa Barbara MRSEC is a member of the Materials Research Facilities Network (www.mrfn.com). Use was made of the computational facilities administered by the Center for Scientific Computing at the CNSI and MRL (an NSF MRSEC; DMR-2308708) and purchased through NSF CNS-1725797. This work was also supported by the National Science Foundation (NSF) through Enabling Quantum Leap: Convergent Accelerated Discovery Foundries for Quantum Materials Science, Engineering and Information (Q-AMASE-i): Quantum Foundry at UC Santa Barbara (DMR-1906325). This research used resources of the Advanced Photon Source, a U.S. Department of Energy (DOE) Office of Science user facility operated for the DOE Office of Science by Argonne National Laboratory under Contract No. DE-AC02-06CH11357.
\end{acknowledgement}

\small
\bibliography{biblio}

\newpage

\centering
\begin{figure}
\includegraphics[width=3.3in]{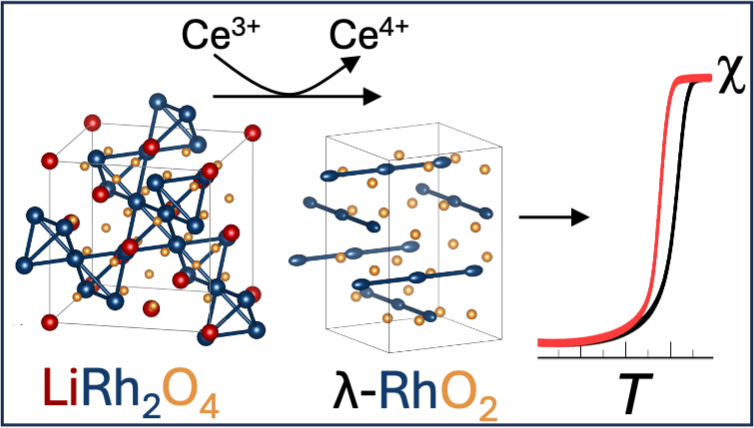}
    \caption*{Table of contents graphic.}
\end{figure}

\end{document}